\documentclass[fleqn,twoside]{article}
\usepackage{espcrc2}

\usepackage{graphicx}
\usepackage[figuresright]{rotating}

\newcommand{\AmS}{{\protect\the\textfont2
  A\kern-.1667em\lower.5ex\hbox{M}\kern-.125emS}}

\title{Weak lensing: Dark Matter, Dark Energy and Dark Gravity}

\author{Alan Heavens\address[IfA]{Scottish Universities Physics Alliance (SUPA), Institute for Astronomy, University of Edinburgh, Blackford Hill, Edinburgh EH9 3HJ, U.K.}}
       
\begin{document}

\begin{abstract}
In this non-specialist review I look at how weak lensing can provide information on the dark sector of the Universe.   The review concentrates on what can be learned about Dark Matter, Dark Energy and Dark Gravity, and why. On Dark Matter, results on the confrontation of theoretical profiles with observation are reviewed, and measurements of neutrino masses discussed.  On Dark Energy, the interest is whether this could be Einstein's cosmological constant, and prospects for high-precision studies of the equation of state are considered.   On Dark Gravity, we consider the exciting prospects for future weak lensing surveys to distinguish General Relativity from extra-dimensional or other gravity theories.
\vspace{1pc}
\end{abstract}

% typeset front matter (including abstract)
\maketitle

\section{Introduction}

Weak lensing (WL) is an attractive probe of structure in the Universe, as it depends only on the distribution of mass; it is blind to the nature of the mass, and is independent of the dynamical state of the matter.  Some uncertainties in interpretation inherent in other methods are consequently removed, and a more direct confrontation of theory and experiment is possible.   Weak gravitational lensing on a large scale, or cosmic shear, is challenging technically, but advances have been rapid as specially-designed instrumentation and methods have been developed.  Indeed, the first cosmic shear detections were only published in the year 2000, and the size of surveys has advanced from $1 \rightarrow 100 \rightarrow 10^4$ square degrees (past, present, near future).  The current state-of-the-art survey is the Canada-France-Hawaii Legacy Survey (CFHTLS), covering $\sim 170$ square degrees, and the precision of cosmological information from weak lensing surveys is now broadly competitive with other probes.  The Pan-STARRS survey ($3\pi$ steradians) starts in mid-2009 and ambitious very deep and wide imaging surveys (Euclid, JDEM, LSST; generically referred to as `future WL surveys' here) are planned, promising very precise investigation of Dark Matter, Dark Energy and Dark Gravity.

\section{Weak lensing}

Weak lensing causes small distortions in the shapes, sizes and apparent brightnesses of distant sources, due to the deflecting influence of non-uniform matter along the line of sight.  Most work has focussed on the change in shape (shear), as precise measurement of sizes is difficult to achieve, and observables arising from changing brightness generally have poorer signal-to-noise than shear.   Shape measurement is usually characterised by a galaxy {\em ellipticity}, which can be defined even if the image is not elliptical, and which is a complex number with a standard deviation typically of 0.3-0.4.  The effect of lensing by general large-scale structure on this is small - around 0.01, so many galaxies are required in order to detect a cosmic shear signal with high signal-to-noise.  Cosmic shear leads to correlations of the ellipticities of galaxy images, and these can be used either statistically, to probe the detailed statistical properties of mass fluctuations in the Universe, or can be used to measure the mass distribution of discrete lensing systems.   Originally mass mapping was done in projection, giving estimates of the surface mass density profile, but now with lensing surveys typically being undertaken with many broad-band filters, distance information through photometric redshifts (photo-zs) can be used as well.  This allows 3D mass mapping to be done, and also gives improved statistical analysis.  The main challenges to realising the promise of future WL surveys are systematic errors (e.g.\cite{Kitching08sys,AmaraRef08,Kitching08sys2}), principally shape measurement errors, biases in photometric redshifts and the shear-intrinsic alignment of background images and foreground galaxies.  Recent advances in shape measurement with {\em lensfit} \cite{lensfit1,lensfit2} mean that this will probably not be a dominant error for CFHTLS and Pan-STARRS, although more development is needed for future surveys.  The photo-z problem is not a fundamental one, but requires a large, deep spectroscopic survey to calibrate.  The last one is, along with uncertainty in the highly nonlinear power spectrum, the dominant theoretical uncertainty, but understanding of these is improving steadily \cite{Hey06,JS08,SchneiderBridle09}.   For more details on weak lensing see, for example \cite{BartelmannSchneider01,vanWaerbekeMellier03,SAASFEE,HJ08,Munshi08}.

\section{Dark Matter}

\begin{figure}[h]
\centering
\includegraphics[width=2.787in]{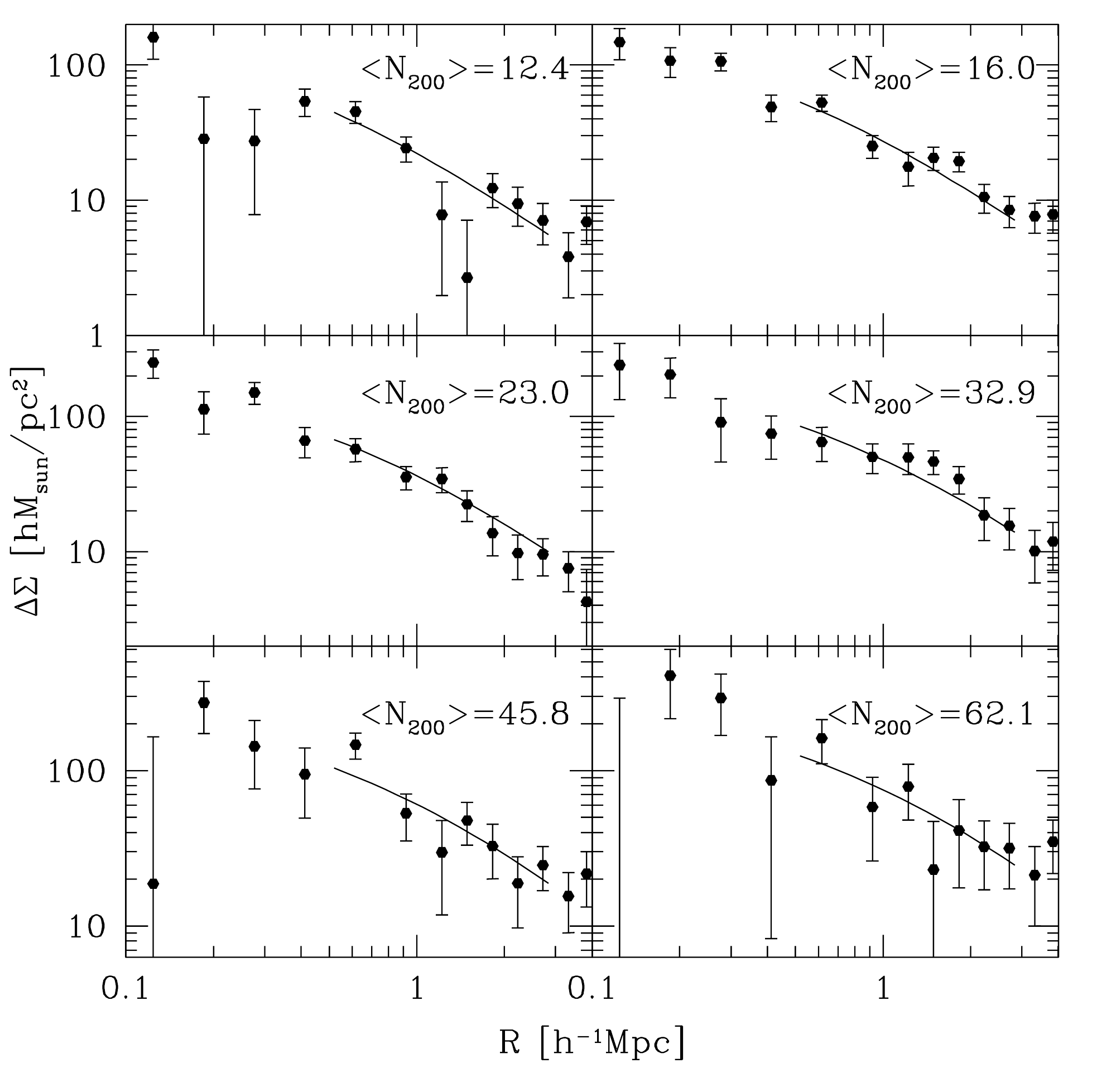}
\caption{Excess surface density from stacked galaxy clusters from the SDSS survey, with best-fitting NFW profiles.  $N_{200}$ is a measure of the richness of the clusters.  From \cite{Mandelbaum08}.
\label{SDSSNFW}}
\end{figure}

Theoretical work with numerical simulations indicates that in the absence of the effects of baryons, virialised Dark Matter haloes should follow a uniform `NFW' profile, $\rho(r) = \rho_s (r_s/r)(1+r/r_s)^{-2}$ \cite{NFW}, if the Dark Matter is cold (CDM).  Simulations also predict how the physical size of the clusters should depend on mass, characterised by the concentration index $c_s \equiv r_{vir}/r_s$, where $r_{vir}$ is defined as the radius within which the mean density is 200 times the background density.  Roughly $c_s \propto M^{-0.1}$. This can be tested, by measuring the shear signal and stacking the results from many haloes to increase signal-to-noise.  Fig. \ref{SDSSNFW} shows the average radial surface density profiles for clusters identified in the Sloan Digital Sky Survey (SDSS), grouped by number of cluster galaxies, and NFW fits superimposed  \cite{Mandelbaum08}.  
\begin{figure}[h]
\centering
\includegraphics[width=2.787in]{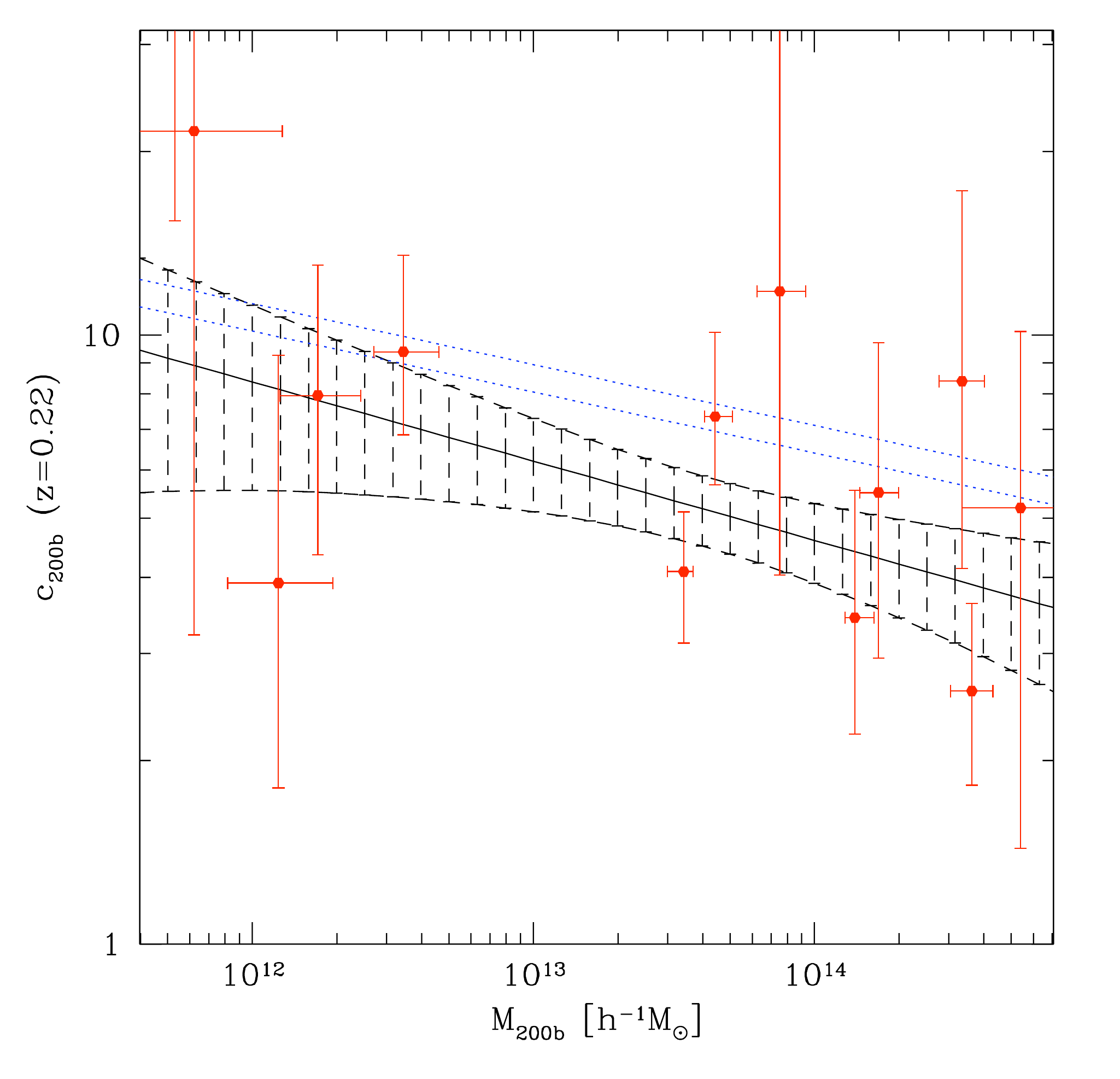}
\caption{Concentration indices from SDSS clusters as a function of mass, compared with simulation (dotted, for different cosmologies).  Dashed regions show range assuming a power-law $c_s-M$ relation.  From \cite{Mandelbaum08}.
\label{SDSSCM}}
\end{figure}
Fig.\ref{SDSSCM} shows that the observed concentration indices are close to the theoretical predictions, but some tension exists.    Broadly, weak lensing data on clusters therefore supports the CDM model.  What about 3D mapping?  This was first done in the COMBO-17 survey by \cite{Taylor04}, and more recently for the COSMOS HST survey \cite{Massey07}.  Unfortunately, the limited accuracy of photo-zs ($\sim 0.03 \simeq 100 h^{-1}$ Mpc typically) means the 3D mass map is smoothed heavily in the radial direction, and this limits the usefulness of 3D mapping for testing the NFW profile.  

A more radical test of theory has been performed with the Bullet Cluster \cite{Clowe04}, actually a pair of clusters which have recently passed through each other.  There are two clear peaks in the surface density of galaxies, and X-ray emission from hot shocked gas in between.  In the standard cosmological model, this makes perfect sense, as the galaxies in the clusters are essentially collisionless.  If the (dominant) Dark Matter is also collisionless, then we would expect to see surface mass concentrations at the locations of the optical galaxy clusters, and this is exactly what is observed.  In MOND or TeVeS models without Dark Matter, one would expect the surface mass density to peak
where the dominant baryon component is - the X-ray gas.  This is not seen.  A caveat is that it is not quite the surface density which is observed, rather the {\em convergence}, which is related to the distortion pattern of the galaxy images, and which is proportional to surface density in General Relativity (GR), but not in MOND/TeVeS.  However, no satisfactory explanation of the bullet cluster has been demonstrated without Dark Matter.

\subsection{Neutrinos}

Since the weak lensing signal depends on the level and evolution of the matter clustering, it is sensitive to neutrino masses through their effect on the matter power spectrum.  The main effect is a suppression of the power on small scales, as the neutrinos free-stream out of fluctuations, and this depends essentially on the sum of the neutrino masses.  Results have been reported for the CFHTLS by \cite{Tereno09}, putting a limit of 3.3 eV (95\%) on the mass sum, reducing to 0.54 eV after including ancillary data.  Prospects for future surveys are very promising, however, with errors of 0.07 eV expected for future surveys \cite{Kitching08nu,Hannestad06}.  This improves on the limits possible from Planck CMB observations by a factor of 4, and is a factor 3 lower than particle physics experiments are expected to be able to achieve by that time \cite{Scranton09}.  Some limited sensitivity to individual neutrino masses is possible \cite{Slosar,Fogli}, (de Bernardis et al., in preparation).

\section{Dark Energy}

The apparent acceleration of the Universe is perhaps most simply accounted for by accepting Einstein's original modification to GR and allowing a non-zero Cosmological Constant.  As such, it is a feature of the law of gravity, but there are other possibilities.  A straightforward modification is to place Einstein's extra term on the other side of the equations, where it acts as a source term for gravity, and corresponds to a vacuum energy density.  This opens up a further possibility, that the source is not a density associated with the vacuum, but rather a new field, Dark Energy, whose energy density may evolve with time.  The Dark Energy needs to have an equation of state parameter $w(a)\equiv p/(\rho c^2)<-1/3$ to drive acceleration.  $w=-1$ corresponds to the Cosmological Constant.  Clearly, a demonstration that $w=-1$ is ruled out by data would have far-reaching consequences for our understanding of the Universe.   Dark Energy can be probed in a number of ways.  Even if it does not cluster, it has measurable effects on weak lensing, which can therefore be used to probe this sector.  There are two main effects.  The first is that Dark Energy modifies the distance-redshift relation $r(z) = c\int_0^z dz'/H(z')$, or equivalently the Hubble parameter 
\begin{eqnarray}
H^2(z) &=& H_0^2\left[\Omega_m a^{-3}+\Omega_k a^{-2}+\right. \nonumber\\
& & \left. \Omega_{DE}
\exp\left(3\int_1^a\,\frac{da'}{a'}\left[1+w(a')\right]\right)\right],
\end{eqnarray}
where $a=(1+z)^{-1}$ is the cosmic scale factor, and $\Omega_{m,k,DE}$ are the current matter, curvature and Dark Energy density parameters.   The Dark Energy also affects the growth rate via the Hubble parameter, since in GR, the fractional overdensity $\delta \equiv \delta \rho/\bar \rho-1$ (where $\bar\rho$ is the mean density) grows to linear order according to
\begin{equation}
\ddot\delta+2 H \dot\delta -4\pi G \rho_m \delta = 0
\end{equation}
where $\rho_m$ is the matter density and we assume the Dark Energy density is not perturbed. 

There are several ways in which weak lensing can be used to probe the Dark Energy.  One is to measure the ratio of average tangential shear as a function of redshift behind clusters.  Theory indicates that this ratio depends on the distance-redshift relation only, with the detailed properties of the mass distribution in the cluster cancelling out. This statistic can be applied to the relatively large signal expected behind galaxy clusters \cite{JainTaylor03,Taylor07}.  In common with other methods, such as study of the luminosity distance of Type Ia supernovae (SN; e.g. \cite{Riess}) or the positions of features in baryonic
acoustic oscillations (BAO; e.g. \cite{EisensteinHu97}), it probes only the expansion
history.

 A complementary approach is to consider the entire observed shear field, and treating it as a sparsely-sampled, noisy 3D shear field.  This approach \cite{Heavens03,Castro05,Heavens06,Kitching07COMBO} retains all of the information in the shear field, and is theoretically very powerful.  It probes the growth rate as well as the geometry, and has the advantage that one can test the gravity model, which we consider later.

A recent study of weak lensing in the CFHTLS gives $-1.18 < w < -0.88$ at 95\% \cite{Kilbinger09}, and next-generation WL surveys should give percent-level errors on the current value of $w$, and errors of around 0.1 on $dw/da$ \cite{Heavens06}.

\section{Dark Gravity}

In addition to the possibility that Dark Energy or the cosmological constant drives acceleration, there is an even more radical solution.  As a cosmological constant, Einstein's term represents a modification of the gravity law, so it is interesting to consider whether the acceleration may be telling us about a failure of GR.  Although no compelling theory currently exists, suggestions include modifications arising from extra dimensions, as might be expected from string-theory braneworld models.  Interestingly, there are potentially measurable effects of such exotic gravity models which weak lensing can probe, and finding evidence for extra dimensions would of course signal a radical departure from our conventional view of the Universe.

\begin{figure}
\centering
\includegraphics[width=2.787in]{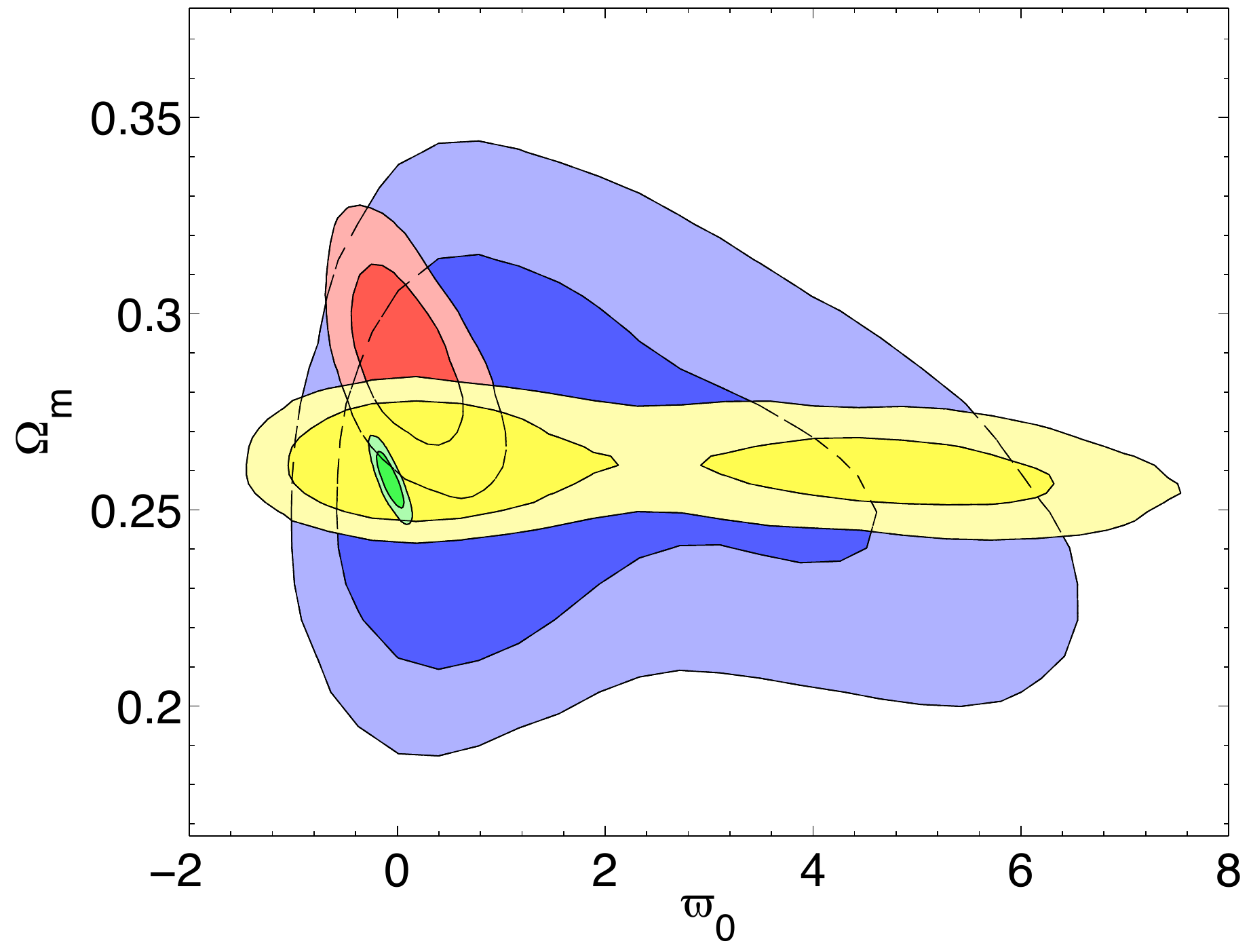}
\caption{The projected marginal 68\% and 95\% likelihood contours for the slip, $\varpi$, assuming $\varpi=\varpi_0(1+z)^{-3}$, for WMAP 5-year data (blue), adding current weak lensing and ISW data (red).  Yellow is mock Planck CMB data, and green adds weak lensing from a 20,000 square degree survey\cite{Daniel}. 
\label{Daniel}}
\end{figure}

If we consider scalar perturbations in the conformal Newtonian gauge (flat for simplicity), $ds^2 = a^2(\eta)\left[(1+2\psi)d\eta^2 - (1-2\phi)d\vec x^2\right]$, where $\psi$ is the potential fluctuation, and $\phi$ the curvature perturbation, $\eta$ being the conformal time.  Information on the gravity law is manifested in these two potentials.  For example in GR and in the absence of anisotropic stresses (a good approximation for epochs when photon and neutrino streaming are unimportant) $\phi=\psi$.   More generally,  the Poisson law may be modified, and the laws for $\psi$ and $\phi$ may differ.  This difference can be characterized\cite{Daniel} by the {\em slip}, $\varpi$.  This may be scale- and time-dependent: $\psi(k,a) = \left[1+\varpi(k,a)\right]\phi(k,a)$,
and the modified Poisson equation may be characterised by  $Q$, an effective change in $G$ \cite{Amendola}:
\begin{equation}
-k^2\phi = 4\pi G a^2 \rho_m \delta_m Q(k,a).
\end{equation}
Different observables are sensitive to $\psi$ and $\phi$ in different ways \cite{JainZhang}.  For example, the Integrated Sachs-Wolfe effect depends on $\dot\psi + \dot\phi$, but the effect is confined to large scales and cosmic variance precludes accurate use for testing modified gravity. Peculiar velocities are sourced by $\psi$.  Lensing is sensitive to $\psi+\phi$, and this is the most promising route for next-generation surveys to probe beyond-Einstein gravity.    The Poisson-like equation for $\psi+\phi$ is
\begin{equation}
-k^2(\psi+\phi) = 2\Sigma\frac{3H_0^2 \Omega_m}{2a} \delta_m
\end{equation}
where $\Sigma \equiv Q(1+\varpi/2)$.   For GR, $\Sigma=1$, $\varpi=0$.  The DGP braneworld model \cite{DGP} has $\Sigma=1$, so mass perturbations deflect light in the same way as GR, but the growth rate of the fluctuations differs.  Thus we have a number of possible observational tests, including probing the expansion history, the growth rate of fluctuations and the mass density-light bending relation.  Future WL surveys can put precise constraints on $\Sigma$ \cite{Amendola}, and on $\varpi$ (see Fig. \ref{Daniel})) \cite{Daniel}.
By probing the growth rate as well as the expansion history, weak lensing can lift a degeneracy which exists in methods which consider the distance-redshift relation alone, since the expansion history in a modified gravity model can always be mimicked by GR and Dark Energy with a suitable $w(a)$.
In general however the growth history of cosmological structures
will be different in the two cases (e.g. \cite{Knox2006,HutererLinder07}, but
see \cite{Kunz}).

\subsection{Growth rate}
Whilst not the most general, the  growth index $\gamma$ \cite{Linder05} is a convenient minimal extension of GR.   
The growth rate of perturbations in the matter density
$\rho_m$, $\delta_m \equiv \delta \rho_m/\rho_m$, is parametrised as a function of scale factor $a(t)$ by
\begin{equation}
\frac{\delta_m}{a} \equiv g(a) =
\exp\left\{\int_0^a\,\frac{da'}{a'}\left[\Omega_m(a')^\gamma-1\right]\right\},
\end{equation}
In the standard GR cosmological model, $\gamma\simeq 0.55$,  whereas in modified gravity 
theories it deviates from this value.  E.g. the flat DGP braneworld model \cite{DGP} has $\gamma\simeq 0.68$
on scales much smaller than those where cosmological 
acceleration is apparent \cite{LinderCahn07}. 
\begin{figure}[h]
\centering
\includegraphics[width=2.787in]{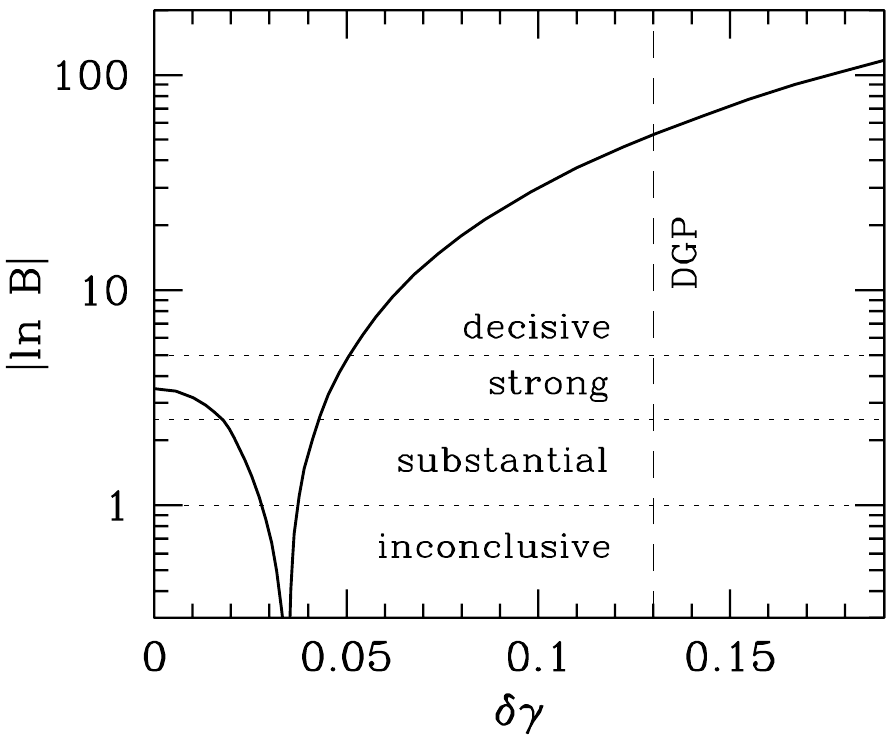}
\caption{Expected Bayesian evidence $B$ vs.
deviation of the growth index from GR, for a future WL survey + \emph{Planck} \cite{Heavens07}.
If modified gravity is the true model, GR will still be favoured by the data to the left of the cusp.
The Jeffreys scale of evidence \cite{Jeffreys61} is labeled.
\label{dgamma}}
\end{figure}
Measurements of the growth factor  can in principle be used to determine the growth index $\gamma$, and it is interesting to know if it is of any practical use.  In
contrast to parameter estimation, this is an issue of 
model selection - is the gravity model GR, or is there evidence for beyond-Einstein gravity?  This question may 
be answered with the Bayesian evidence, $B$ \cite{Skilling04}, which is the ratio of 
probabilities of two or more models, given some data.
Following  \cite{Heavens07}, Fig.  \ref{dgamma} shows how the Bayesian evidence for GR changes 
with increasing true deviation of $\gamma$ from its GR value 
for a combination of a future WL survey and \emph{Planck}. 
From the WL data alone, one should be able to distinguish GR decisively 
from the flat DGP model at $\ln B \simeq 11.8$, or, 
in the frequentist view, $5.4\sigma$ \cite{Heavens07}.
The combination of WL+\emph{Planck}+BAO +SN should be able to distinguish $\delta \gamma = 0.041$ at $3.41$ sigma. This data  combination should be able to decisively distinguish a Dark Energy GR model
from a DGP modified-gravity model with expected
evidence ratio $\ln B \simeq 50$.  An alternative approach is to explore whether the expansion history and growth rate are consistent, assuming GR\cite{Ishak,Song}.

One caveat on these conclusions is that WL requires knowledge of nonlinear clustering, and this is reasonably well-understood for GR, but for other models, further theoretical work is needed. This has already started\cite{Schmidt2008}.


\begin{thebibliography}{9}

\bibitem{AmaraRef08}
Amara A., Refregier A., 2008, MNRAS, 391, 228

\bibitem{BartelmannSchneider01}
Bartelmann M., Schneider P., 2001, Phys. Reports, 340,291

\bibitem{Amendola}Amendola L., Kunz M., Sapone D., 2008, JCAP, 04, 13A

\bibitem{Castro05} Castro, P. G.; Heavens, A. F.; Kitching, T. D.; 2005, PhRvD, 72, 3516
 
\bibitem{Clowe04}Clowe D., Gonzalez A., Markevitch M.,  2004, ApJ, 604, 596
 
\bibitem{Daniel}Daniel S., Caldwell R., Cooray A., Serra P., Melchiorri A., astroph/0901.0919

\bibitem{DGP}
{Dvali} G., {Gabadadze} G., and {Porrati} M., Phys. Lett. B 
{\bf 485},  208  (2000).
 
\bibitem%[Eisenstein \& Hu (1997)]
{EisensteinHu97}Eisenstein D. J.,  Hu W., 1997, ApJ, 511, 5

%\bibitem%[Fisher (1935)]
%{Fisher}Fisher R.A., 1935, J. Roy. Stat. Soc., 98, 39

\bibitem{Fogli}
Fogli G.L. et al., 2008, Phys. Rev. D78, 033010

\bibitem{Heavens03} Heavens, A. F., 2003, MNRAS, 343, 1327

\bibitem{Heavens06} Heavens, A. F., Kitching, T. D., Taylor, A. N., 2006, MNRAS, 373, 105

\bibitem{Heavens07}
{Heavens} A.F., {Kitching} T.D., {Verde} L., 2007, MNRAS, 380, 1029
 
\bibitem{Hannestad06} Hannestad S., Tu H., Wong Y., 2006, JCAP 0606, 025

\bibitem{Hey06}
Heymans C., White M., Heavens A., Vale C., van Waerbeke L., MNRAS, 371, 750

%\bibitem%[Hobson, Bridle \& Lahav (2002) ]
%{Hobson} Hobson M.P., Bridle S.L., Lahav O., 2002, MNRAS, 335, 377

\bibitem{HJ08}
Hoekstra H., Jain B., 2008, ARNPS, 58, 99

\bibitem%[Huterer \& Linder (2007)]
{HutererLinder07} Huterer D., Linder E.V., 2007, PRD, 75, 2, 3519

\bibitem{Ishak}Ishak M., Upadhye A., Spergel D., 2006, Phys. Rev. D., 74, 3513

\bibitem{JainTaylor03}Jain B., Taylor A.N., PRL, 91, 141302 (2003)

\bibitem{JainZhang}Jain B., Zhang P., Phys. Rev. D, 2008, 78, 3503

\bibitem{Jeffreys61}
H. {Jeffreys}, {\em {Theory of Probability}} 
(Oxford University Press, UK, 1961).

\bibitem{JS08}
Joachimi B., Schneider P., 2008, A\& A, 488, 829

\bibitem{Kilbinger09}
Kilbinger M., et al., 2009, A\& A, 497, 677

\bibitem{Kitching07COMBO} Kitching, T. D. et al., 2007, MNRAS, 376, 771

\bibitem{Kitching08sys} Kitching, T. D., Taylor, A. N., Heavens, A. F.,  2008,  MNRAS,  389, 173

\bibitem{Kitching08sys2}
Kitching T.D., Amara A., Abdalla F.B., Joachimi B., Refregier A., 2008, astroph 0812.1966

\bibitem{lensfit2}
Kitching T.D., Miller L., Heymans C.E., van Waerbeke L., Heavens A.F., 2008, MNRAS, 390, 149

\bibitem{Kitching08nu} Kitching T. D., Heavens A. F., Verde L., Serra P., Melchiorri A., 2008, PRD, 77, 10, 103008

\bibitem{Knox2006}Knox L., Song Y.-S., Tyson J.A., 2006, Phys. Rev. D, 74, 3512
 
\bibitem%[Kunz \& Sapone (2007)]
{Kunz} Kunz M., Sapone D., 2007, PRL, 98, 12, 121301

%\bibitem{linder05b}
%{Linder} E. V., Astropart. Phys. {\bf 24},  391  (2005).

\bibitem{Linder05}
{Linder} E. V., PRD{} {\bf 72},  043529  (2005).

\bibitem{LinderCahn07}
{Linder} E. V., {Cahn} R.N., Astropart. Phys. {\bf 28},  481  (2007).

%\bibitem%[Liddle et al. (2006)]
%{LMPW} Liddle A., Mukherjee P., Parkinson D., Wang Y., 2006, PRD, 74, 13, 12506

\bibitem{Mandelbaum08}
Mandelbaum R., Seljak U., Hirata C.M., JCAP 8, 6 (2008).

\bibitem{Massey07}Massey R., et al.,  2007, Nature, 445, 286

\bibitem{lensfit1}
Miller L., Kitching T., Heymans C., Heavens A.F., van Waerbeke L., 2007, MNRAS, 382, 315

\bibitem{Munshi08}
Munshi D., Valageas P., van Waerbeke L., Heavens A.F., 2008, Phys. Rev., 462, 67 

\bibitem{NFW}
Navarro J., Frenk C.S, White S.D.M., 1997, ApJ, 490, 493.

\bibitem%[Riess et al. (1998)]
{Riess} Riess A., et al., 1998, AJ, 116, 1009

\bibitem{Scranton09}
Scranton R., et al, 2009, astroph/0902.2590

\bibitem{Schmidt2008}Schmidt F., Lima M., Oyaizu H., Hu W., 2008, astroph/0812.0545

\bibitem{SchneiderBridle09}
Schneider M., Bridle S., 2009, astroph/0903.3870

\bibitem {SAASFEE} 
Schneider P., Kochanek C., Wambsganss J.
 {\em Gravitational lensing: Strong, Weak and Micro}, 2006, Saas-Fee advanced course 33, XVI, 552, 196

%\bibitem%[Serra, Heavens \& Melchiorri (2007)]
%{Serra07} Serra P., Heavens A.F., Melchiorri A., 2007, MNRAS, 379, 169

\bibitem%[Skilling (2004)]
{Skilling04} Skilling J., 2004, avaliable at\\
 http://www.inference.phy.cam.ac.uk/bayesys

\bibitem{Slosar}
Slosar A., 2006, Phys. Rev. D73, 123501
 
\bibitem{Song}Song Y.-S., Dor\' e O., 2009, JCAP, 1208, 039

%\bibitem%[Szydlowski \& Godlowski (2006a)]
%{Szydlowski06a} Szydlowski M., Godlowski W., 2006a, Phys. Lett. B633, 427
%\bibitem%[Szydlowski \& Godlowski (2006b)]
%{Szydlowski06b} Szydlowski M., Godlowski W., 2006b, Phys. Lett. B639, 5

%\bibitem{TBHT}Taylor A.N., Ballinger W.E., Heavens A.F., Tadros H., 2001, MNRAS, 327, 689

\bibitem{Taylor04}Taylor A.N. et al.,  2004, MNRAS, 353, 1176

\bibitem%[Taylor et al. (2006)]
{Taylor07} Taylor A.N., Kitching T.D., Bacon D.J., Heavens A.F., 2007, MNRAS, 374, 1377

\bibitem{Tereno09}
{Tereno} I., {Schimd} C., {Uzan} J.-P., {Kilbinger}, M., {Vincent} F.~H., {Fu} L., 2008, astroph 0810.0555

\bibitem{vanWaerbekeMellier03}van Waerbeke L., Mellier Y., {\em Gravitational Lensing by Large Scale Structures: A
Review}. Proceedings of Aussois Winter School, astroph/0305089
(2003)

%\bibitem{WSP2008}White M., Song Y.-S., Percival W.J., 2008, astroph/0810.1518

\end{thebibliography}
\end{document}